\documentclass[superscriptaddress,unsortedaddress,twocolumn,floatfix]{revtex4}
\usepackage{textcomp} 
\usepackage{graphicx}

\begin{document}

\title{Preparation and magnetoresistance of  Ag$_{2+x}$Se thin films deposited via Pulsed Laser Deposition.}
\author{B. Mogwitz}
	\affiliation{Physikalisch-Chemisches Institut, Justus-Liebig-Universit\"{a}t,\\ Heinrich-Buff-Ring 58, 35392 Gie\ss{}en, Germany}
\author{C. Korte}
	\affiliation{Physikalisch-Chemisches Institut, Justus-Liebig-Universit\"{a}t,\\  Heinrich-Buff-Ring 58, 35392 Gie\ss{}en, Germany}
\author{M. v. Kreutzbruck}
	\affiliation{Institut f\"{u}r Angewandte Physik Justus-Liebig-Universit\"{a}t,\\  Heinrich-Buff-Ring 16, 35392 Gie\ss{}en, Germany}
\author{L. Kienle}
	\affiliation{Max-Planck-Institut f\"{u}r Festk\"{o}rperforschung,\\ Heisenberg Str. 1, 70569 Stuttgart,  Germany}
\author{J. Janek}
	\email{Juergen.Janek@phys.chemie.uni-giessen.de}
	\affiliation{Physikalisch-Chemisches Institut, Justus-Liebig-Universit\"{a}t,\\  Heinrich-Buff-Ring 58, 35392 Gie\ss{}en, Germany}
\begin{abstract}
The preparation of Ag$_{2+x}$Se thin films with thicknesses between 4 nm and 3 \textmu{}m by pulsed laser deposition 
on single crystalline NaCl and MgO substrates is reported.  
The films are perfectly dense and show a good lateral uniformity with a small number of defects. The microstructure of the films corresponds to a nanoparquet, being composed of two different phases of silver selenide. One phase is identified as the Naumannite low temperature phase of silver selenide, the structure of the other phase has not been reported in detail before and probably represents a metastable phase. Silver-rich films contain silver precipitates with typical sizes on the nanoscale. Their presence and their size appears to be responsible for the large and linear magnetoresistance effect of silver-rich silver selenide.
\end{abstract}


\maketitle

\section{Introduction}
The mixed-conducting nonstoichiometric silver and copper chalcogenides Ag$_{2+x}$X and 
Cu$_{2+x}$X (X = S, Se, Te) are well known for their unusual transport 
properties. Thus, a number of phases exhibit both a high electronic and a high 
ionic conductivity, leading to extremely high chemical diffusion coefficients 
in those phases which also possess a high thermodynamic factor \cite{Miyatani57,Miyatani58}. In addition, 
silver selenide and silver telluride are currently studied as interesting high 
temperature MR (magneto resistance) materials \cite{Xu97,Husmann02,Chuprakov98,Manoharan01}.
The preparation of well defined bulk material Ag$_{2+x}$X and  Cu$_{2+x}$X is mostly based on the 
co-melting of metal and chalcogen or on reactive growth from the elements \cite{Ohachi92}.
Unfortunately both systems undergo phase transformations, and  it is a  difficult task to grow
untwinned single crystals of the room temperature phases. In the present case of Ag$_{2+x}$Se a phase transformation from the 
low temperature $\alpha$-phase  to the high temperature $\beta$-phase occurs at $T = 406~K$. Thus, Ag$_{2+x}$Se being prepared by co-melting is always polycrystalline.
An additional problem is caused by the nonstoichiometry $x$. Despite being small, it controls the
defect concentrations,  and thus, may have a significant influence on the transport properties \cite{Kreutzbruck05}.
Obviously, the magnitude of this direct influence depends on the intrinsic defect concentrations. 
The low and
high temperature phases  are characterized by maximum deviations from
stoichiometry  $x$ of $10^{-4}$ and $10^{-3}$, respectively.
As the band gap of Ag$_{2+x}$Se is very small ($\Delta E_g \simeq 0.1 eV$), such small metal excess creates
only small changes of the electronic conductivity, however, the concentration of ionic defects increases considerably. This may change the electronic conductivity indirectly, by increasing the concentration of scattering centers.

$x$ can only be fixed to a desired value by equilibration of a Ag$_{2+x}$Se 
sample with a silver or selenium reservoir with known activity or by coulometric titration.
Studies of the MR effect of $\alpha$-Ag$_{2+x}$Se (low-temperature phase) with such well controlled $x$ have only been reported by
\cite{Beck04}, and show that the size of the MR effect indeed depends strongly on~$x$.

Mostly driven by the diverse and attractive MR properties of Ag$_{2+x}$Se, the need for reproducible thin film preparation arises.
But to the best of our knowledge there is yet no method available to grow Ag$_{2+x}$Se films of high quality, i.e.
with well defined morphology, composition and thickness. To explore the best method for the 
deposition of Ag$_{2+x}$Se, we therefore investigated different methods for the preparation of thin films of silver
and copper chalcogenides, focusing to silver selenide as the most prominent magnetoresistive
material within the group of materials. As the chemical
properties of silver and copper  chalcogenides are quite close, the general results are
probably valid for all of them. Basically we compared three different approaches: (a) reactive growth
of thin films starting from thin films of metal and elemental chalcogen, (b) thermal
evaporation of the compounds, (c) pulsed laser deposition (PLD) of the compounds.
Results from the first two methods are already described in the literature \cite{Das89} \cite{Safran92}, and our
findings are close to those. So the major purpose of this paper is to present the results of 
the PLD and to discuss its advantages and disadvantages compared to other methods.

The paper is organized as follows: Firstly, we
briefly describe the preparation of thin Ag$_{2+x}$Se films by pulsed laser deposition in argon as background gas.
Secondly, we summarize the structural, microstructural and morphological properties of the thin films, 
as the microstructure, morphology and the metal to chalcogen ratio (and its gradient across a film) are the key factors 
for the quality of a film. Thirdly, we report on MR effect measurements
of the films prepared by PLD.

Finally, we compare our findings with previous reports on the MR effect in bulk and thin film material and show that
PLD provides thin films with better quality than those grown by PVD (physical vapor deposition), sputtering or reactive growth.

\section{Experimental}

\textit{PLD}: For the deposition process of Ag$_{2+x}$Se we used a vacuum chamber (minimum pressure $p=5 \cdot 10^{-3}$ Pa) equipped with 
a turbomolecular pump and a cold trap (liquid nitrogen) to avoid a selenium contamination of the pumping system.
A KrF-excimer laser 
($\lambda =$ 248 nm, Compex 201, Lambda-Physik, G\"o{}ttingen, Germany) was used for the ablation process. In order to optimize the film quality with respect to 
homogeneity and density we varied the pulse energy $E_{pulse}$ between 200 and 250 mJ (approximately (6~-~8)~$\frac{J}{cm^2}$), the repetition rate $\nu_{rep}$ between 5 and 10 Hz,
the pressure of the background gas between 1 and 10 Pa and the substrate temperature between 308 and 473 K.
Argon was used as the background gas. 
The distance between the target and the substrate can be varied between 40 and 
100~mm, but in all experiments shown here the distance was  adjusted to 45~mm.
Films with a thickness between 4~nm and 3~\textmu{}m were prepared. 

\textit{Characterisation}: The morphology of all films was routinely characterised by HRSEM (high resolution scanning electron microscopy, LEO Gemini 982) and the composition was determined with EDX (energy dispersive X-ray spectroscopy, Oxford INCA).
Selected samples were investigated by XRD (x-ray diffractometry) and in situ XRD (2$\Theta-\Theta$ geometry). Sufficiently thin
films were placed on alumina grids which were fixed in a side-entry,
double-tilt holder. High resolution transmission electron microscopy (HRTEM)
and selected area electron diffraction (SAED) were performed in a Philips
CM30ST (300~kV, LaB$_6$ cathode, Cs = 1.15~mm). EDX  was performed in the scanning- and nanoprobe mode of CM30ST
with a Si/Li-EDX detector (Noran, Vantage System). 

\textit{Targets:}
Bulk silver selenide as target material for the laser ablation was prepared by
co-melting of the elements in sealed silica ampoules at 1200 K (for 24 h, left to cool in air) in the appropriate ratio.   
During the ablation the target rotated continuously, and the point of incidence of the laser was shifted in order to
obtain a laterally more homogeneous ablation. Targets with different composition were used to prepare films with well known
silver/selenium ratio.

\textit{Substrates:}
We used freshly cleaved MgO single crystal pieces (10x10 mm$^2$, 1 mm thickness) with well reproducible adhesive properties.
Freshly cleaved single crystalline NaCl was used for the preparation of TEM samples. In this case we deposited thin films with a thickness of only 20 to 30~nm. 
These silver selenide films were easily separated from the NaCl substrate in water (cf. \cite{Safran98}) and then transfered to TEM aluminium grids.
Copper grids which are more common in TEM could not be used, as they undergo a displacement reaction with silver selenide even at room temperature. 
For some of the XRD measurements we also used fused (x-ray amorphous) silica as substrate, being cleaned
with hot chromic/sulfuric acid before the deposition. 

\textit{Temperature:}
The substrate temperature during the deposition was measured in the substrate holder (steel)
at the back of the the target with a Ni/CrNi-thermocouple. An additional thermocouple fixed on the surface 
of the substrate was used to check the temperature during the deposition. The temperature difference between
substrate surface and the back of the substrate never exceeded 4 to 5~K.

\textit{Magnetoresistance-(MR)-effect:}
The MR effect of all films were measured by a standard four-probe dc method. The magnetic field was applied by a 
17~T superconducting magnet system (Oxford) with the temperature control ranging from 2 to 300~K.

\textit{Film thickness:}
After the MR measurements the substrate/film samples were cleaved to allow a precise 
thickness measurement by HRSEM. For film thicknesses below 25~nm the resolution of the HRSEM was not sufficient, and we calculated 
the thickness from the deposition time (extrapolated with the almost linear relation between film thickness and deposition time).

\section{Experimental results}

\subsection{Characterisation by HRSEM and EDX}
\begin{figure}
 \includegraphics[scale=0.3]{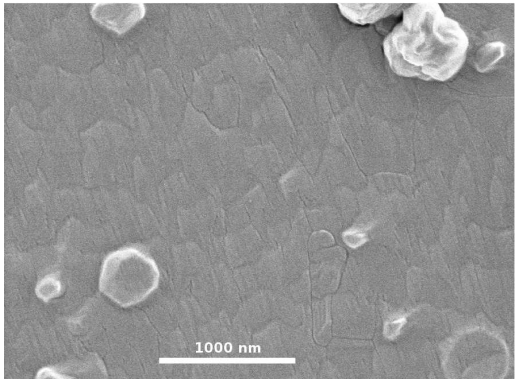}
	\caption{\label{fig:V0703D} Surface of a Ag$_{2+x}$Se film (500~nm thickness prepared by PLD ($p=2$~Pa (Ar), 
	$T = 383$~K, $\nu_{rep} = 10$~Hz, $E_{pulse} = 200$~mJ); HRSEM micrograph}
	\end{figure}

\begin{figure}
 \includegraphics[scale=0.3]{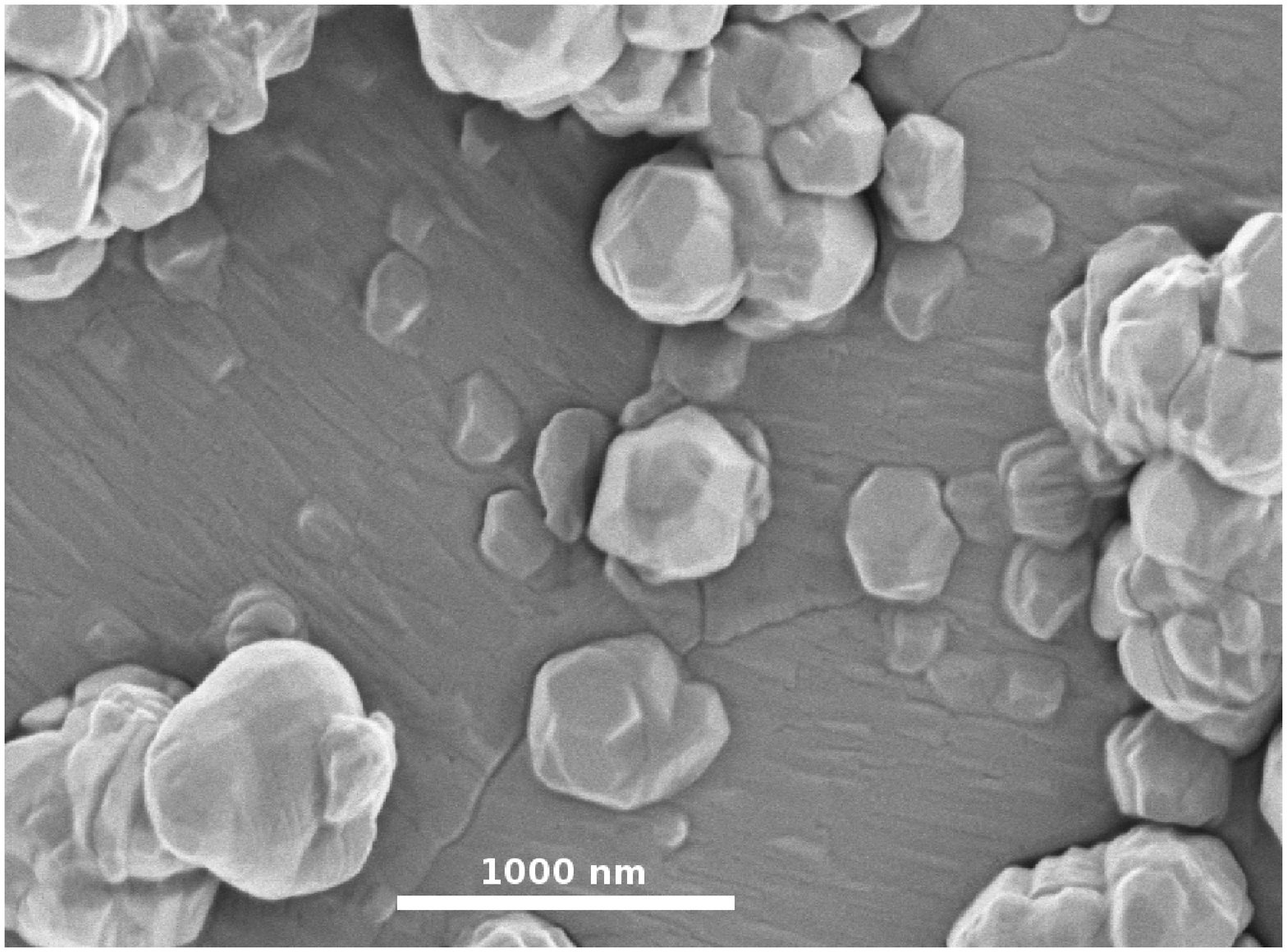}
	\caption{\label{fig:V1203D} Surface of a Ag$_{2+x}$Se film (500~nm thickness prepared by PLD ($p=10$~Pa (Ar), 
	$T = 383$~K, $\nu_{rep} = 10$~Hz, $E_{pulse} = 200$~mJ); HRSEM micrograph}
	\end{figure}

\begin{figure}
 \includegraphics[scale=0.3]{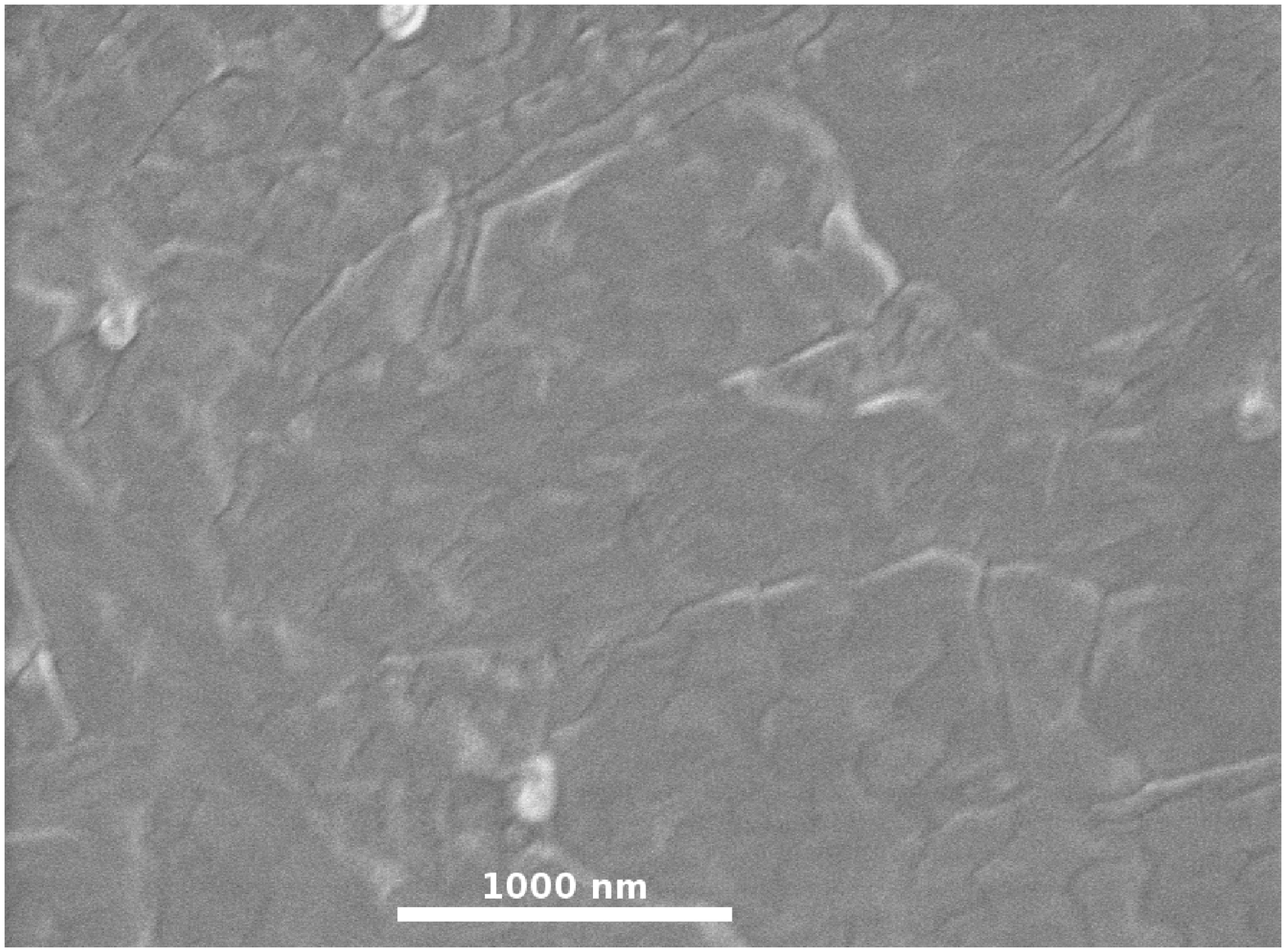}
	\caption{\label{fig:V1303B} Surface of a Ag$_{2+x}$Se film (600~nm thickness prepared by PLD ($p=2$~Pa (Ar), 
	$T = 383$~K, $\nu_{rep} = 10$~Hz, $E_{pulse} = 200$~mJ); HRSEM micrograph}
	\end{figure}

\begin{figure}
 \includegraphics[scale=0.3]{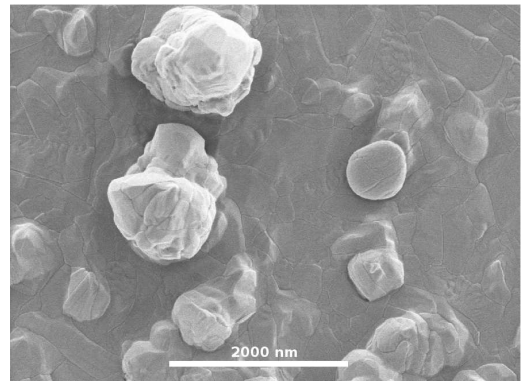}
	\caption{\label{fig:V1803B} Surface of a Ag$_{2+x}$Se film (700~nm thickness prepared by PLD ($p=2$~Pa (Ar), 
	$T = 383$~K, $\nu_{rep} = 7$~Hz, $E_{pulse} = 200$~mJ); HRSEM micrograph}
	\end{figure}

Figs. \ref{fig:V0703D} and \ref{fig:V1203D} show two representative surfaces of films which have been deposited under different argon pressure, and  
figs. \ref{fig:V1303B} and \ref{fig:V1803B} show typical surface areas of films which have been deposited with a different repetition rate.
Both higher argon background pressures and smaller repetition rate led to a poor film quality with many large droplets and crystallites on
top of the surface.
\\
The background pressure influences also the selenium concentration in the film. With increased pressure and therefore higher flow rate of the background gas
we found  a decrease of the selenium concentration from (30.1$\pm$0.3) atom-\% to (28.8$\pm$0.4) atom-\% by EDX. Regions or areas with an increased silver concentration could not be detected, i.e. the films are homogeneous with respect
to the chemical composition.

\begin{figure} 
 \includegraphics[scale=0.63]{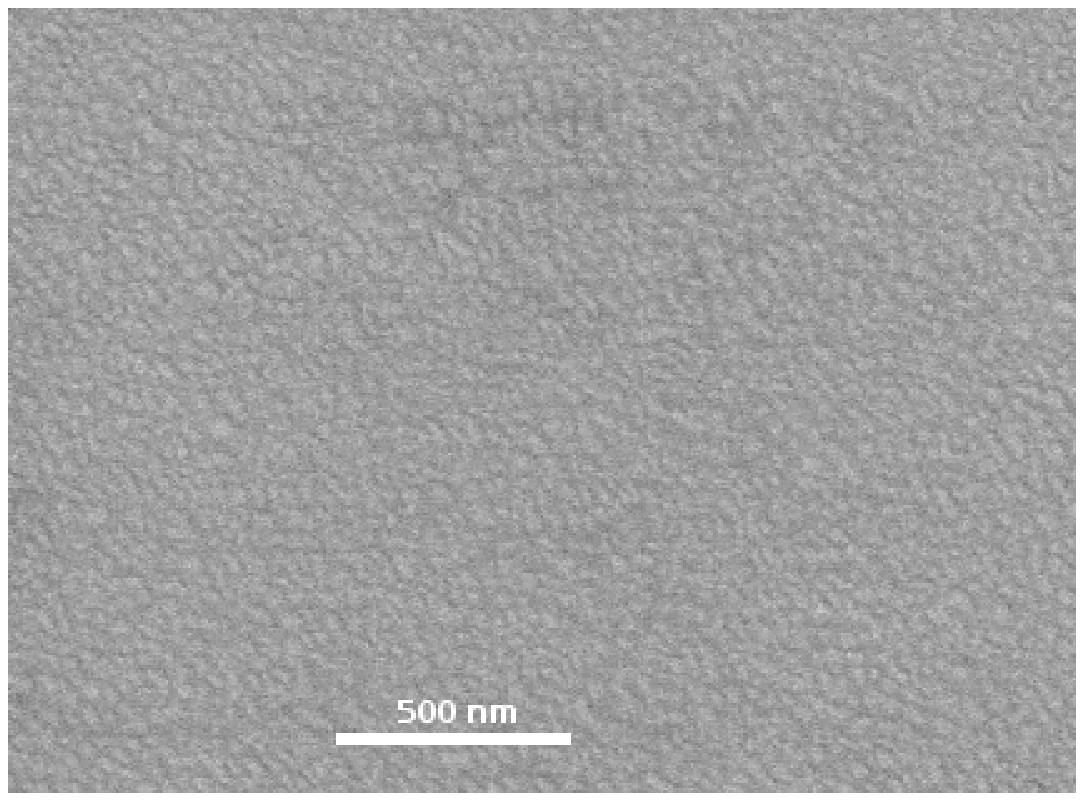}
	\caption{\label{fig:V0810B} Surface of a Ag$_{2+x}$Se film (5~nm thickness prepared by PLD ($p=2$~Pa (Ar), 
	$T = 383$~K, $\nu_{rep} = 10$~Hz, $E_{pulse} = 200$~mJ); HRSEM micrograph}
	\end{figure}

\begin{figure}
 \includegraphics[scale=0.49]{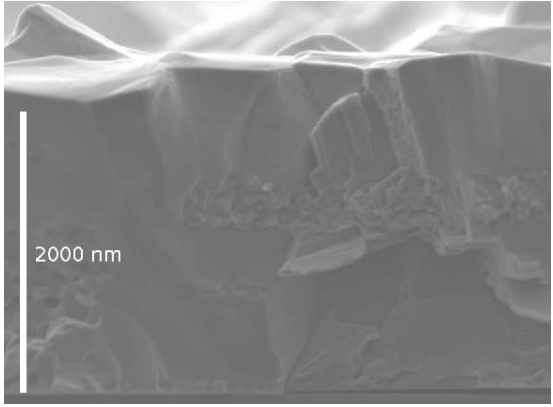}
	\caption{\label{fig:V0810E} Cross section (not polished) of a Ag$_{2+x}$Se film (2050~nm thickness prepared by 
	PLD ($p=2$~Pa (Ar),  $T = 383$~K, $\nu_{rep} = 10$~Hz, $E_{pulse} = 200$~mJ); HRSEM micrograph}
	\end{figure}

Films being prepared under optimized conditions show a very flat and regular surface morphology with only very few defects (pits). Fig.~\ref{fig:V0810B} shows the surface of a 5~nm thick film with a flat and virtually defect-free surface.
The cross section of a film with a thickness of about 2 \textmu{}m in fig.~\ref{fig:V0810E} proves that the film interior is usually free of pores and polycrystalline with
a columnar structure. 

A systematic variation of the deposition parameters led to the result that a background pressure of about 2~Pa Argon with a minimum flow rate and a laser repetition rate of 10~Hz at a power of 200~mJ 
is the optimum for the preparation of high quality silver selenide films with high lateral uniformity. 

\subsection{Transmission electron microscopy}
\begin{figure}
 \includegraphics[scale=0.8]{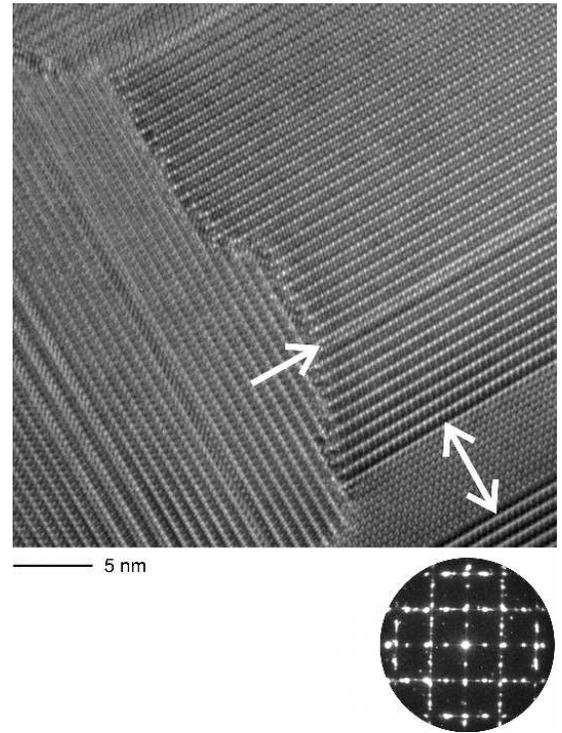}
	    \caption{\label{fig:tem1}  Parquet-like real structure of $\alpha$-Ag$_{2+x}$Se film with SAED pattern. The
			arrows highlight one thick and one single layer lamella of TT2, see text.
	    }
	\end{figure}

\begin{figure}
 \includegraphics[scale=0.9]{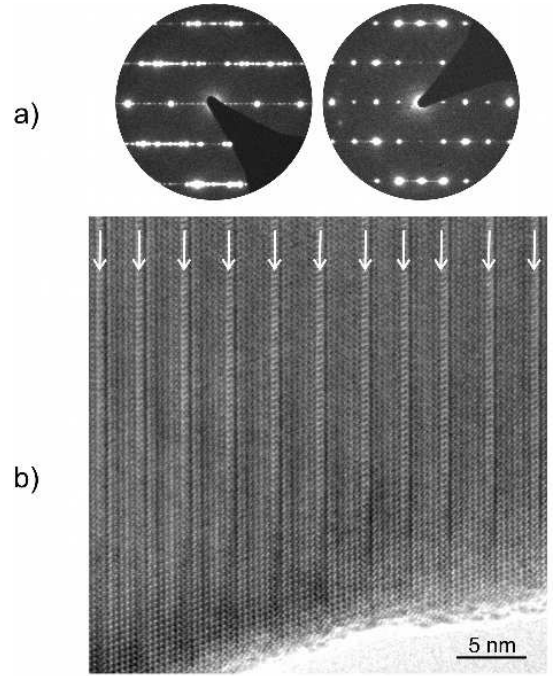}
	    \caption{\label{fig:tem2}a) SAED patterns recorded in distinct selected areas of the same
		film, [010]$_{TT1}$. b) HRTEM micrograph with almost periodic sequence of single
		layer defects into a thick TT1-type layer.
	    }
	\end{figure}

The microstructure of the samples is characterized by a rectangular
parquet-like arrangement of lamellar domains, cf. fig.~\ref{fig:tem1}. SAED patterns of
wide areas of the films represent the superposition of all lamellar
components, hence, pseudo-fourfold symmetry is produced, see diffraction
pattern in fig.~\ref{fig:tem1}. The rectangular arrangement of the domains is based on
twinning, see discussion in ~\cite{Janek04}. As evidenced by EDX analyses
significant fluctuations of the (Ag : Se) ratio are not present, and the
average ratio is close to the expected value of 2 : 1. By focusing the
electron beam on thicker lamellas with the same orientation a
Naumannite-type phase (TT1~\cite{Wiegers71}) can be identified as the major
component. The minor component can be assigned to a new modification of
Ag$_{2+x}$Se (TT2) which was described in ~\cite{Janek04}. Both modifications
show a fully coherent intergrowth. As a rule, TT2 forms very thin lamellas
with an average thickness of a few nanometers (cf. double-headed arrow in
fig.~\ref{fig:tem1}), however, occasionally single layers of TT2 in a matrix of TT1 can
be demonstrated, see arrow in fig.~\ref{fig:tem1}. Single layer defects are rarely
observed and their arrangement is random thus producing faint diffuse
streaks in the SAED patterns of thick TT1 lamellas, see fig.~\ref{fig:tem2}a, left. In
a few selected areas of the films, the diffuse intensity is concentrating in
broad reflections, see fig.~\ref{fig:tem2}a right. In this case, the HRTEM micrographs
exhibit an almost periodic sequence of single layer defects with nearly
equal distances between them. The existence of nanostructures with single
layer defects again underlines the high compatibility of layers with TT1-
and TT2-type structure.

\subsection{In situ X-ray diffractometry}
\begin{figure}
 \includegraphics[scale=0.7]{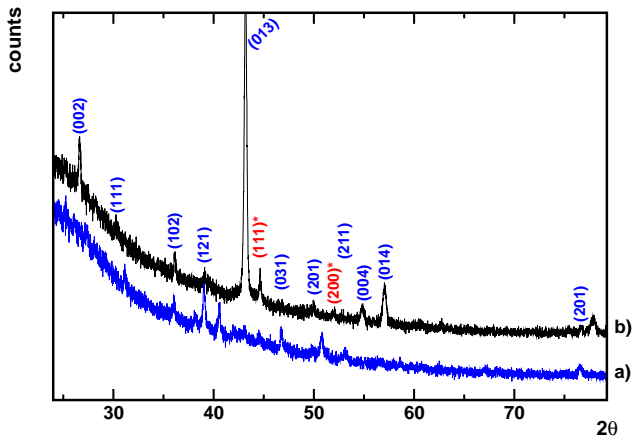}
	    \caption{\label{fig:V150104-2} XRD of a  Ag$_{2+x}$Se thin film (low temperature $\alpha$-phase); $170$~nm thickness; Co-K$_\alpha$; $T=303$~K\\
	    	a) Film as deposited at 303~K; b) Film  after a double phasetransformation
		 $\alpha\rightarrow\beta\rightarrow\alpha$.
		Reflections of silver metal are marked with an asterisk. All other reflections can be attributed to the 
		orthorhombic Naumannite low temperature phase.
	    }
	\end{figure}

\begin{figure}
 \includegraphics[scale=0.34]{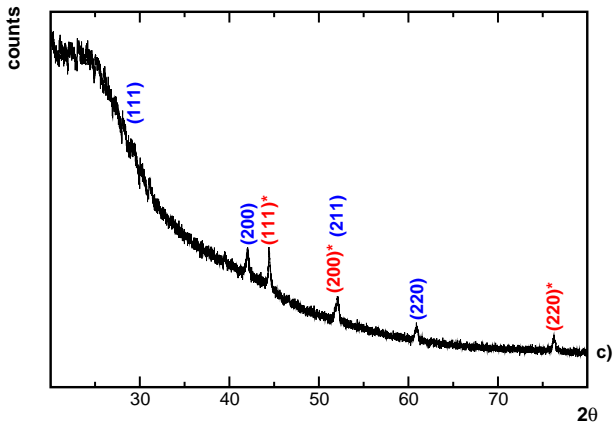}
	    \caption{\label{fig:V150104-3} XRD of a Ag$_{2+x}$Se thin film (high temperature $\beta$-phase); $170$~nm thickness; Co-K$_\alpha$; $T=473$~K
	    	(deposited at 303~K).
		Reflections of silver metal are marked with an asterisk;
		All other reflections can be attributed to the cubic high temperature phase.
	    }
	\end{figure}

In fig. \ref{fig:V150104-2} and \ref{fig:V150104-3} the results of an in situ XRD experiment are summarized. The temperature of the film (deposited on fused silica at about $300$~K) 
was increased from 303 to 503~K and then decreased back to room temperature. 
The phase transformation from the orthorhombic low temperature phase to the cubic high temperature phase between 
403 and 413~K could well be 
identified. After the heating/cooling cycle the reflections of the low temperature phase (TT1) appear more pronounced, i.e. the microstructure of the film
has changed during the cycle. The highly diffuse background is caused by the fused silica substrate (Herasil 102, supplied by Heraeus, Germany).

We always found the silver (111) reflection in the as-deposited films, as exemplied in fig. \ref{fig:V150104-2}. 
This clearly indicates that the volume fraction of silver metal is at least of the order of 1\%, as otherwise the silver reflex would not be visible. This is in good agreement with a maximum volume fraction of silver in the order of 10\% for $x\approx0.5$.	
The reflection is small and broadened, due to
the small size of the silver precipitates. Upon heating from room temperature the FWHM (full width at half-maximum)
reduces, indicating ripening and growth of the silver precipitates. The Ag (111) reflection is also found after the phase
transformation to the cubic phase above 406~K. After cooling back to room temperature the Ag (111) reflection has a significantly
smaller FWHM. 

\begin{figure}
 \includegraphics[clip,scale=0.34]{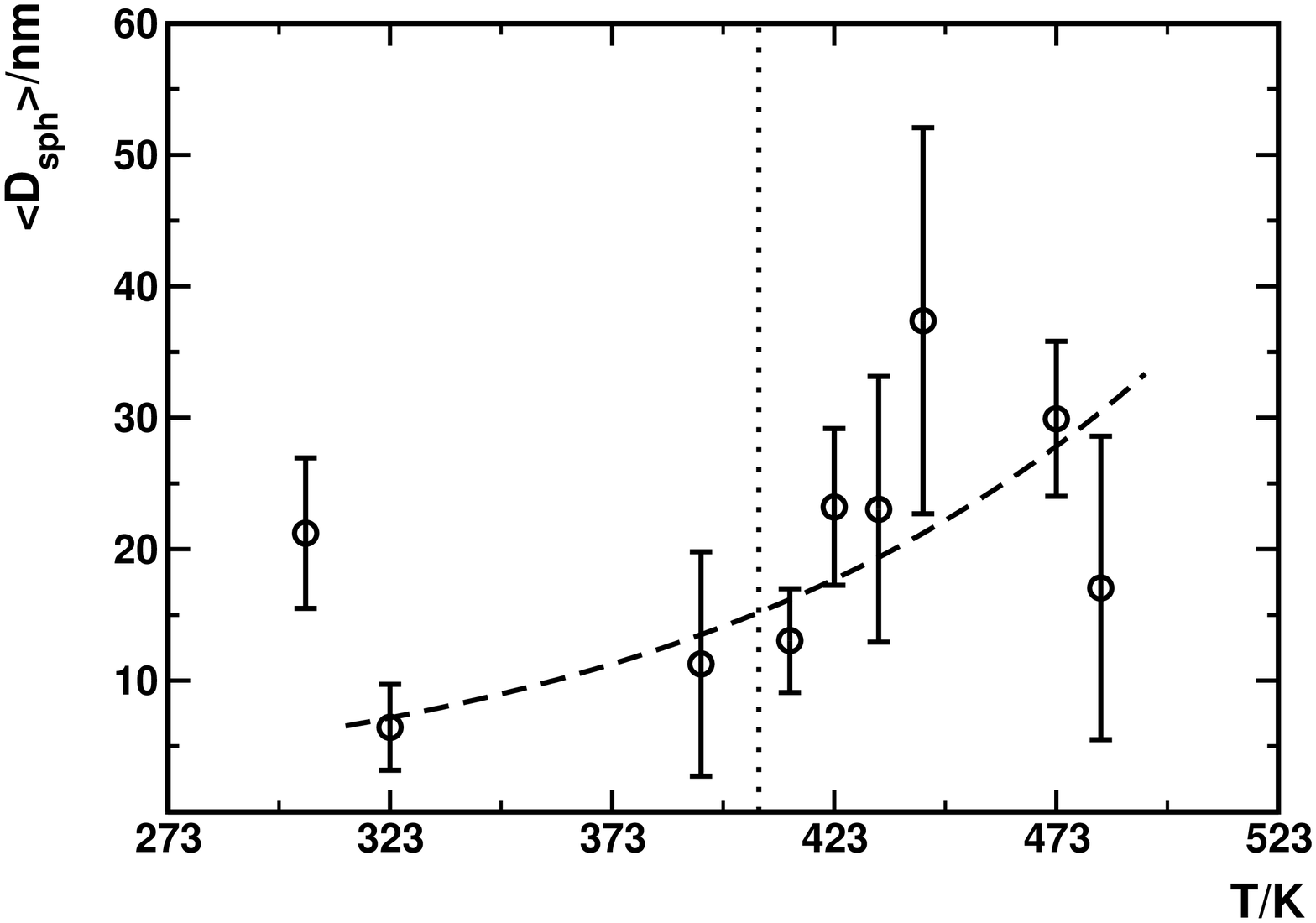} 
	    \caption{\label{fig:V14Scherrer} Diameter of Ag precipitates in Ag$_{2+x}$Se  at different temperatures according to the Scherrer equation. The point at 303~K refers to the size after one heating cooling cycle.
	    }
	\end{figure}

We fitted the (111) reflection profile with a convoluted Cauchy Gauss distribution function (Voigt) \cite{root5}, and
using the Scherrer equation. The instrument broadening was determined using Si powder as standard. 
Under the assumption that the silver metal forms spherical-shaped crystallites \cite{Birkholz06} we calculated the silver particle size for 170 nm films  ($x=0.54\pm{}0.04$), see fig. \ref{fig:V14Scherrer}.  The data scatter considerably, but nevertheless offer reasonable quantitative and qualitative information.
The size of the particles for this film ranges  from 7~nm before to 21~nm after heating above the transition temperature. Thus, silver excess in the form of very small nanoparticles is only obtained at lower deposition temperatures (300~K).
   
\printfigures

\subsection{MR effect}
\begin{figure}
 \includegraphics[clip,scale=0.34]{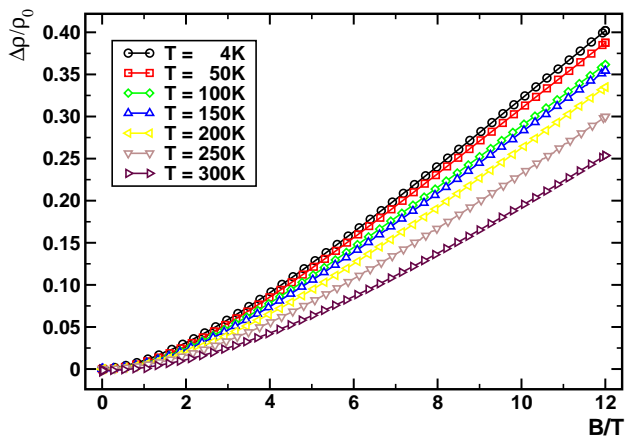}
	    \caption{\label{fig:V1203-MR} MR effect for the Ag$_{2+x}$Se thin film shown in fig. \ref{fig:V1203D}.}
	\end{figure}

\begin{figure}
 \includegraphics[clip,scale=0.34]{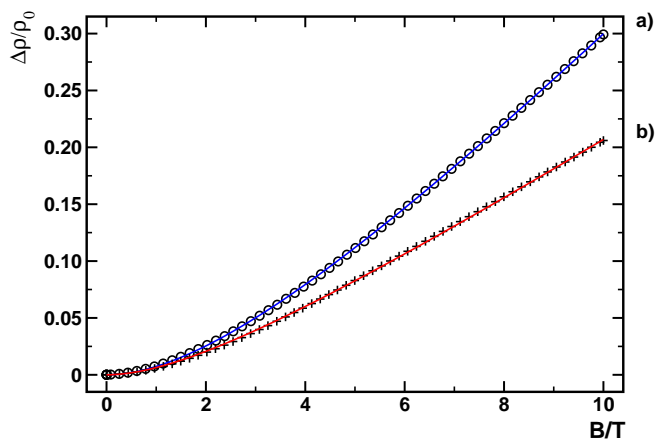}
	    \caption{\label{fig:V0512-MR} MR effect measured at 4~K for a Ag$_{2+x}$Se thin film; thickness about 2~\textmu{}m) . 
	    a) As deposited (at $\approx$ 300~K);
	    b) After 3~h annealing at 433~K}.
	\end{figure}

The  transversal MR effect vs. the magnetic flux up to 12~T is depicted in fig.\ \ref{fig:V1203-MR} for a film with~500 nm thickness and a composition (measured with EDX) of Ag$_{2+x}$Se  ($x=0.47\pm{}0.05$, deposited at room temperature). The field dependence and the size of the MR effect is close to the results for bulk material as reported by Manoharan \cite{Manoharan01}. Up to a magnetic field of about 2 to 3~T we found a quadratic dependence which changes to a linear dependence at higher magnetic fluxes. The measurements are well fitted with the empirical equation
$MR = \frac{aB^2}{(b+cB)}$.
The MR effect does not saturate even at 12 T.
As shown in fig.\ \ref{fig:V0512-MR} the MR effect decreased after annealing.  XRD
measurements before and after the heat treatment (3~h at 433~K) show a change in the film microstructure as well as
an increased size of the silver precipitates in this case from $(131\pm13)$ to $(237\pm52)$~nm (diameter for spherical particles) according to the Scherrer equation.

\section{Discussion}

The films prepared by PLD show a high quality regarding their density, their compositional homogeneity and their morphology. A high background pressure reduces the free mean path of ions, atoms and small molecular clusters during PLD and increases the deposition rate of larger clusters and droplets. In the present case an argon pressure of 2~Pa with a small flow rate led to the best results. A higher repetition rate of the laser pulses reduces the time for the ripening of surface defects, whereas a small repetition rate reduces the growth rate of the film. With a repetition rate of 10~Hz we obtained the best results, i.e. we obtained dense films with high lateral uniformity and a small concentration of droplets. 

As we found by HRTEM the silver selenide films prepared by PLD have a specific and unusual microstructure, characterised by the appearance of a second phase of Ag$_{2+x}$Se (TT2) in coexistence with the well known Naumannite phase (TT1). This microstructure is best described as a nano-parquet with relatively thick lamellae of the Naumannite phase and relatively thin lamellae of the TT2 phase. Thus, PLD films of silver selenide are always nanostructured and contain a high concentration of twin boundaries. The phase diagram of the Ag-Se system contains only TT1 as the thermodynamically stable low temperature phase. Thus, we assume that TT2 is a metastable phase being a result of the PLD process. To the best of our knowledge, the TT2 phase has not been observed before in films being prepared by sputtering or PVD.

With respect to a large and linear MR effect, the silver excess plays a decisive role. A clear and unequivocal explanation of the MR properties of silver-rich silver selenide in terms of a specific micro- and defect structure is still missing, but we found that the presence of very small silver precipitates is a necessary prerequisite for a linear MR effect. In this respect, PLD is unique as it offers the possibility to deposit films under non-equilibrium conditions with a virtually homogeneous chemical composition. The results of the in situ XRD indicate that as deposited films contain excess silver in the form of very small precipitates with mean diameters below 10~nm. Upon heating from room temperature to higher temperatures the size of the silver precipitates increases. Along with the increase of the particle size the MR effect decreases, as to be seen in fig. \ref{fig:V0512-MR}. Thus, we conclude that a large linear MR effect relies on the presence of silver precipitates with a very small size. 
In a previous work we found that a silver dispersion on the atomic scale does not maximize the MR effect, rather it leads to a negative MR effect \cite{Beck01}. Thus, we suggest that there exists an optimal microstructure of the silver/silver selenide dispersion with silver being dispersed in the form of nanosized clusters or nuclei. However, these rather small particles have a strong tendency to ripen, and we have to assume that ripening does even occur at room temperature, due to the large chemical diffusion coefficient of silver in the silver selenide matrix at room temperature or above.
This degradation will become slower at larger sizes of the Ag-clusters, but it will not stop, inevitably resulting in a reduction of the MR-effect. Therefore the silver/silver selenide nanodispersion will be a suitable sensor material with only when it comes to low temperature applications. The problem of sufficient long term stabilities needs to be addressed especially when using very small silver precipitates with sizes of a few nm, which are far from equilibrium condition.
With these findings the way to systematic studies of the correlation between microstructure, chemical composition and MR effect in silver selenide is open, and further measurements are underway.

\section{Conclusions}

PLD offers an excellent route for the preparation of very thin (about 10 nm) and relatively thick (\textmu{}m range) and dense silver selenide films
with different composition. The films are homogeneous with respect to the silver/selenium ratio, i.e. the fast ablation of target material by the excimer laser
minimize the preferential vaporisation of selenium. The deposition in a wide range of temperatures allows to prepare films with different microstructures. The possibility of the deposition of meta stable films at lower temperature with respect of the MR-Effect seems especially promising.
\\
The deposition parameters background pressure, substrate temperature, laser power and repetition rate	can be controlled such that either almost stoichiometric 
or silver-rich films are deposited. Droplet formation cannot be suppressed completely, but appears not to be a problem for the reproducibility of the MR effect.

The main advantage of the PLD in comparison to other methods is the possibility to produce 
homogeneous films in a broad range of thicknesses. We can avoid lateral or transversal 
silver concentration gradients as seen in films prepared by PVD methods. 
The MR effect is well reproducible for the same film, but scatters for different films with the same thickness,
which is due to variations of the precipitate size and their spatial distribution.
Due to the continuous
process it is also possible to study the film during the growth or the cooling down in situ by various 
methods (for example conductivity measurement).

\begin{acknowledgments}
This study was funded by the DFG (project Ja648/5-1). Financial support by the FCI is also gratefully acknowledged (JJ). We are grateful to Dr.~Lakshmi (Institute of Physical Chemistry, RWTH Aachen) for the in situ XRD measurements. We thank F.~Gruhl and G.~Lembke for help with the MR effect measurements. 
\end{acknowledgments}

\newpage 

%\bibliography{PLD-paper}

\newpage

\end{document}